\documentclass{article}
\usepackage{amssymb}
\usepackage{xypic}
\xyoption{all}
\newtheorem{definition}{Definition}[section]

\newtheorem{corollary}[definition]{Corollary}

\newtheorem{lemma}[definition]{Lemma}

\newenvironment{proof}{\noindent {\bf Proof. }}{\hspace{\fill}$\Box$}

\newcommand{\id}{\mbox{\rm id}}
\title{On Reversible Cellular Automata with Triplet
Local Rules}
\author{
Shuichi INOKUCHI\thanks{Faculty of Mathematics, Kyushu University 33,
Fukuoka 812-8581, Japan.}, \
Kazumasa HONDA\thanks{Department of Informatics, Kyushu University 33,
Fukuoka 812-8581, Japan.}, \
Hyen Yeal LEE\thanks{School of Computer Engineering,
Pusan National University, Pusan 609-735,
Korea.}, \\
Tatsuro SATO\thanks{Oita National College of Technology,
Oita, 870-0152, Japan.}, \
Yoshihiro MIZOGUCHI$^{*}$ and
Yasuo KAWAHARA$^{\dag}$
}
\date{}
\begin{document}
\maketitle
\begin{abstract}
Bijections between sets may be seen as discrete (or
crisp) unitary transformations used in quantum
computations. So discrete quantum cellular automata
are cellular automata with reversible transition
functions. This note studies on 1d
reversible cellular automata with triplet local
rules.
\end{abstract}
\section{Introduction}

Since Feynman proposed the notion of ``quantum
computation'', a lot of models of quantum
computation have been investigated. 
Watrous\cite{watrous-1995} introduces the notion of quantum cellular automata(QCA, for short)
of a kind of quantum computer
and showed that any quantum Turing machines can be simulated by 
a partitioned QCA(PQCA) with constant slowdown.
Moreover he presented necessary and sufficient conditions for the
well-formedness of 1d PQCA.
Watrous' QCA have infinite cell arrays.
Inokuchi and Mizoguchi\cite{inokuchi-2005} introduced 
a notion of cyclic QCA
with finite cell array, which generalises PQCA,
and formulated sufficient condition for local transition functions
to form QCA.
Quantum computations are performed by means of
applying unitary transformations for quantum states.
So classical cellular automata(CA) with reversible transition
functions are considered as a special type of QCA.

On the other hand CA have also been studied
as models of universal computation and complex systems
\cite{wolfram-1986,goles-1990,lee-1996,inokuchi-1996}. 
Reversible CA and the reversibility of CA
were discussed by many researchers.
Wolfram\cite{wolfram-2002} investigated the reversibility of several
models of CA and showed that only six CA,
whose transition functions are identity function, right-shift function,
left-shift function and these complement functions,
of the 256 elementary CA with infinite cell array
are reversible.
Dow\cite{dow-1997} investigated the injectivity(reversibility) of additive CA with
finite cyclic cell array and infinite cell array, and
showed the relation between injectivities of these additive CA.
Morita and Harao\cite{morita-1989} showed that for any reversible Turing
machine there is a reversible CA that simulates it. 

This paper treats with 1d CA, denoted by $CA-R(n)$,
with finite cell array, triplet local rule of rule number $R$
and two states, and investigate the reversibility of $CA-R(n)$.
We can easily observe dynamical behaviors of
$CA-R(n)$ by computer simulations, and consequently
get the following table which shows whether 
1d cellular automata $CA-R(n)$ are reversible or not,
according to five types of boundary conditions
$a$-$b$, $a$-$*$, $*$-$b$, $*$-$*$ and $*$,
which will be defined in the next section.

\begin{center}
\begin{tabular}{|c||c|c|c|c|c|}
\hline
Rule numbers & $a$-$b$ & $a$-$*$ & $*$-$b$ & $*$-$*$ & ~$*$~ \\
\hline
51, 204 & $\bigcirc$ & $\bigcirc$ & $\bigcirc$
& $\bigcirc$ & $\bigcirc$ \\
\hline
15, 85, 170, 240 &&&&& $\bigcirc$ \\
\hline
90, 165 & $\bigcirc$ & $\bigcirc$ & $\bigcirc$
&& \\
\hline
60,195 & $\bigcirc$ & $\bigcirc$ &  &  & \\
\hline
102,153 & $\bigcirc$ & & $\bigcirc$ &  & \\
\hline
150,105 & $\bigcirc$ & $\bigcirc$ & $\bigcirc$
& $\bigcirc$ & $\bigcirc$ \\
\hline
166,180,154,210,89,75,101,45&&&&&$\bigcirc$ \\
\hline
\end{tabular}
\end{center}

Although for 1d CA with infinite cell array 
there exist only six trivial reversible CA, 
we will prove that there exist several non
trivial reversible 1d CA with finite cell array.

\section{Cellular Automata $CA{-}R$}

Cellular automata treated in the paper have
linearly ordered and finite number cells
bearing with states $0$ or $1$. The next state of
any cell depends upon the states of left cell,
the cell itself and the right cell. In this section
we will formally define cellular automata $CA{-}R(n)$
with rule number $R$ of triplet local rule and
$n$ cell array. 

Let $Q$ be a state set $\{0,1\}$ and $n$ a positive
integer. The complement of a state $a\in Q$ will
be denoted by $a^-$, that is, $a^-=1-a$. The state
set $Q$ forms an additive group by the addition $+$
(modulo $2$) (the exclusive logical sum), that is,
$0+0=1+1=0$ and $0+1=1+0=1$. Remark that $a^-=1+a$
for all states $a\in Q$.
The $n$-th cartesian product of $Q$ is denoted by
$Q^n$, in other words, $Q^n$ is the set of all
$n$-tuples consisting of $0$ and $1$. For example,
\[Q^3=\{000,001,010,011,100,101,110,111\}.\]
An $n$-tuple $x=x_1x_2\cdots x_n\in Q^n$ may be
called a word of length $n$ over $Q$, or a
configuration in a context of cellular automata.
It is obvious that the $n$-th cartesian product $Q^n$
also forms an additive group by the component-wise
addition, that is,
\[x_1x_2\cdots x_n+y_1y_2\cdots y_n=
(x_1+y_1)(x_2+y_2)\cdots(x_n+y_n)\]
for all $n$-tuples $x_1x_2\cdots x_n,y_1y_2
\cdots y_n\in Q^n$. \\

A {\it triplet local (transition) rule} is a
function $f:Q^3\to Q$ and the {\it rule number}
$|f|$ of $f$ is defined by
\[\begin{array}{ccll}
|f| &=& \sum_{a,b,c\in Q}2^{4a+2b+c}f(abc). \\
\end{array}\]

\noindent
Note that the rule number $|f|$ is a natural
number with $0\le|f|\le 255$.
A triplet local rule with rule number $R$ will be
denoted by $f_R$, namely $|f_R|=R$. 


Let $f:Q^3\to Q$ be a triplet local rule.
The {\it symmetric rule} $f^{\sharp}:Q^3\to Q$ of $f$
is defined by \[f^{\sharp}(abc)=f(cba)\]
for all triples $abc\in Q^3$. It is trivial that
$f^{\sharp\sharp}=f$ and
\[|f^{\sharp}|=|f|+56(r_3-r_6)+14(r_1-r_4)\]
where $r_{4a+2b+c}=f(abc)$ for all triples
$abc\in Q^3$. 
The {\it complementary rule} $f^-:Q^3\to Q$ of $f$
is defined by \[f^-(abc)=[f(abc)]^-\]
for all triples $abc\in Q^3$.
Note that $f^{-\,-}=f$ and $f^{-\,\sharp}
=f^{\sharp\,-}$ and
\[|f^-|=255-|f|.\]
The {\it reverse rule} $f^\circ:Q^3\to Q$ of $f$
is defined by \[f^\circ(abc)=[f(a^-b^-c^-)]^-\]
for all triples $abc\in Q^3$. It is trivial that
$f^{\circ\circ}=f$, $f^{\sharp\circ}=f^{\circ\sharp}$
and
\[|f^\circ|=255-(128r_0+64r_1+32r_2+16r_3+8r_4
+4r_5+2r_6+r_7).\]

A {\it dynamical system} is a pair $(X,\delta)$
of a set $X$ and a transition function
$\delta:X\to X$. 

Let $f:Q^3\to Q$ be a triplet local rule, $n$
a positive integer and $a,b\in Q$.
By setting different boundary conditions
we can define five transition functions
$f_{(n),a-b}:Q^n\to Q^n$, $f_{(n),*}:Q^n\to Q^n$,
$f_{(n),*-*}:Q^n\to Q^n$, $f_{(n),a-*}:Q^n\to Q^n$
and $f_{(n),*-b}:Q^n\to Q^n$ as follows.

\[\begin{array}{l}
f_{(n),a{-}b}(x_1x_2\cdots x_{n-1}x_n)
=f(a\,x_1x_2)f(x_1x_2x_3)\cdots
f(x_{n-2}x_{n-1}x_n)f(x_{n-1}x_nb), \\
\\
f_{(n),*}(x_1x_2\cdots x_{n-1}x_n)
=f(x_nx_1x_2)f(x_1x_2x_3)\cdots
f(x_{n-2}x_{n-1}x_n)f(x_{n-1}x_nx_1), \\
\\
f_{(n),*{-}*}(x_1x_2\cdots x_{n-1}x_n)
=f(x_1x_1x_2)f(x_1x_2x_3)\cdots
f(x_{n-2}x_{n-1}x_n)f(x_{n-1}x_nx_n), \\
\\
f_{(n),a{-}*}(x_1x_2\cdots x_{n-1}x_n)
=f(a\,x_1x_2)f(x_1x_2x_3)\cdots
f(x_{n-2}x_{n-1}x_n)f(x_{n-1}x_nx_n), \\
\\
f_{(n),*{-}b}(x_1x_2\cdots x_{n-1}x_n)
=f(x_1x_1x_2)f(x_1x_2x_3)\cdots
f(x_{n-2}x_{n-1}x_n)f(x_{n-1}x_nb)
\end{array}\]
for all $n$-tuples $x_1x_2\cdots x_{n-1}x_n\in Q^n$
and all states $a,b\in Q$. 

Set rule number $R=|f|$ ($0\le R\le 255$).
Cellular automata $CA{-}R_{a-b}(n)$ with fixed boundary $a{-}b$,
$CA{-}R_*(n)$ with cyclic boundary,
$CA{-}R_{*-*}(n)$ with free boundary,
$CA{-}R_{a{-}*}(n)$ with right free boundary $a{-}*$ and  
$CA{-}R_{*{-}b}(n)$ with left free boundary $*{-}b$
are dynamical systems $(Q^n,f_{(n),a-b})$, 
$(Q^n,f_{(n),*})$,
$(Q^n,f_{(n),*-*})$,
$(Q^n,f_{(n),a-*})$ and 
$(Q^n,f_{(n),*-b})$, respectively.
This is denoted by the followings for short;
\[CA{-}|f|_{\{a-b,a-*,*-b,*-*,*\}}(n)
=(Q^n,f_{(n),\{a-b,a-*,*-b,*-*,*\}}).\]

\noindent
Transition functions $\delta:Q^n\to Q^n$ are not
always bijections, but when it is the case we can
regard them as discrete quantum automata. 

\begin{definition}
{\rm
\begin{enumerate}
\item A triplet local rule $f:Q^3\to Q$ is
{\it additive} if $f(abc+a'b'c')=f(abc)+f(a'b'c')$
for all triples $abc,a'b'c'\in Q^3$.
\item A transition function $\delta:Q^n\to Q^n$
is {\it additive} if $\delta(x+x')=\delta(x)+\delta(x')$
for all configurations $x,x'\in Q^n$.
\end{enumerate}
}
\end{definition}
We now recall the basic fact on the reversibility
of dynamical systems over $Q^n$.

\begin{lemma}\label{fact-additive}
\begin{enumerate}
\item A transition function $\delta:Q^n\to Q^n$
is bijective iff it is injective iff it is surjective.
\item If a triplet local rule $f:Q^3\to Q$ is
additive, then so is the transition function
$f_{(n),\{0-0,0-*,*-0,*-*.*\}}:Q^n\to Q^n$ for
all positive integers $n$.
\item An additive transition function
$\delta:Q^n\to Q^n$ is bijective iff $\delta(x)=0^n$
implies $x=0^n$ for all configurations $x\in Q^n$.
\end{enumerate}
\end{lemma}
\begin{proof}
 (1) It is trivial since the set $Q^n$
is finite. Also (2) and (3) are clear.
\end{proof}

\section{Basic Results}

In this section we show some general properties
of cellular automata $CA{-}R(n)$. 
Let $(X,\delta)$ and $(Y,\gamma)$ be two dynamical
systems. An {\it isomorphism}
$t:(X,\delta)\to(Y,\gamma)$ is a bijection $t:X\to Y$
rendering the following square commutative:
\[\xymatrix{
X \ar[rr]^t\ar[d]_\delta && Y \ar[d]^\gamma \\
X \ar[rr]_t && Y.
}\]
We call $(X,\delta)$ and $(Y,\gamma)$ isomorphic,
denoted by $(X,\delta)\cong(Y,\gamma)$, if there
exists an isomorphism between $(X,\delta)$ and
$(Y,\gamma)$. It is trivial that isomorphic
dynamical systems are essentially the same ones.

\begin{lemma}\label{fact-symmetric}
The followings holds;
 \begin{enumerate}
  \item $CA{-}|f^{\sharp}|_{\{a-b,a-*,*-b,*-*,*\}}(n)
	\cong CA{-}|f|_{\{b-a,*-a,b-*,*-*,*\}}(n)$.
  \item $CA{-}|f^\circ|_{\{a-b,a-*,*-b,*-*,*\}}(n)
	\cong CA{-}|f|_{\{a^--b^-,a^--*,*-b^-,*-*,*\}}(n)$.
 \end{enumerate}
\end{lemma}
\begin{proof}
 \begin{enumerate}
  \item \label{fact-symmetric-proof}
	This fact asserts the following
	five statements:
	\begin{enumerate}
	 \item $CA{-}|f^{\sharp}|_{a-b}(n)\cong CA{-}|f|_{b-a}(n)$,
	 \item $CA{-}|f^{\sharp}|_{a-*}(n)\cong CA{-}|f|_{*-a}(n)$,
	 \item $CA{-}|f^{\sharp}|_{*-b}(n)\cong CA{-}|f|_{b-*}(n)$,
	 \item $CA{-}|f^{\sharp}|_{*-*}(n)\cong CA{-}|f|_{*-*}(n)$,
	 \item $CA{-}|f^{\sharp}|_*(n)\cong CA{-}|f|_*(n)$.
	\end{enumerate}
	It is easy to show that a bijection $t_n:Q^n\to Q^n$
	defined by $t_n(x_1x_2\cdots x_n)=x_n\cdots x_2x_1$
	gives an isomorphism. We will prove only
	$t_n\circ f_{(n),a-b}=f^{\sharp}_{(n),b-a}\circ t_n$.
	\[\begin{array}{ccll}
	 (t_n\circ f_{(n),a-b})(x_1x_2\cdots x_n)
	  &=& t_n(f(ax_1x_2)f(x_1x_2x_3)\cdots f(x_{n-1}x_nb)) \\
	   &=& f(x_{n-1}x_nb)\cdots f(x_1x_2x_3)f(ax_1x_2) \\
	   &=& f^{\sharp}(bx_nx_{n-1})\cdots f^{\sharp}(x_3x_2x_1)f^{\sharp}(x_2x_1a) \\
	   &=& f^{\sharp}_{(n),b-a}(x_n\cdots x_2x_1) \\
	   &=& (f^{\sharp}_{(n),b-a}\circ t_n)(x_1x_2\cdots x_n).
	  \end{array}\]
  \item 
	This can be shown in the same way as \ref{fact-symmetric-proof}.\\
 \end{enumerate}
\end{proof}
\begin{corollary}\label{corollary-symmetric-reverse}
\[\begin{array}{ccll}
CA{-}|f|_{\{a-b,a-*,*-b,*-*,*\}}(n)
&\cong& CA{-}|f^{\sharp}|_{\{b-a,*-a,b-*,*-*,*\}}(n) \\
&\cong& CA{-}|f^\circ|_{\{a^--b^-,a^--*,*-b^-,*-*,*\}}(n) \\
&\cong& CA{-}|f^{\sharp\,\circ}|_{\{b^--a^-,*-a^-,
b^--*,*-*,*\}}(n).
\end{array}\]
\end{corollary}
Thus a quartet $[|f|,|f^{\sharp}|,|f^\circ|,
|f^{\sharp\,\circ}|]$ of rule numbers $|f|$,
$|f^{\sharp}|$, $|f^\circ|$ and $|f^{\sharp\,\circ}|$
is an equivalence class of rule numbers which
represent isomorphic triplet local rules.
For example, $[102,60,153,195]$ and $[89,75,101,45]$.
\begin{lemma}\label{fact-complement}
~\\
$CA{-}|f^-|_{\{a-b,a-*,*-b,*-*,*\}}(n)$ is reversible
iff so is $CA{-}|f|_{\{a-b,a-*,*-b,*-*,*\}}(n)$.
\end{lemma}
\begin{proof} 
It simply follows from
\[f^-_{(n),\{a-b,a-*,*-b,*-*,*\}}
=c_n\circ f_{(n),\{a-b,a-*,*-b,*-*,*\}},\]
where $c_n:Q^n\to Q^n$ is a bijection defined by
$c_n(x_1x_2\cdots x_n)=x_1^-x_2^-\cdots x_n^-$.
We will prove only $f^-_{(n),a-b}=c_n\circ f_{(n),a-b}$.
\[\begin{array}{ccll}
f^-_{(n),a-b}(x_1x_2\cdots x_n)
&=& f^-(ax_1x_2)f^-(x_1x_2x_3)\cdots f^-(x_{n-1}x_nb) \\
&=& c_n(f(ax_1x_2)f(x_1x_2x_3)\cdots f(x_{n-1}x_nb)) \\
&=& (c_n\circ f_{(n),a-b})(x_1x_2\cdots x_n).
\end{array}\]
\end{proof}
\begin{lemma}
Let $k$ and $n$ be positive integers.
If $CA{-}|f|_*(kn)$ is reversible, then
so is $CA{-}|f|_*(n)$.
\end{lemma}
\begin{proof}
 We will see that the injectivity of
$f_{(kn),*}:Q^{kn}\to Q^{kn}$ implies that of
$f_{(n),*}:Q^n\to Q^n$. The result easily follows
from a fact that the identity
\[\begin{array}{ccll}
&& f_{(kn),*}(x_1x_2\cdots x_nx_1x_2\cdots x_n
\cdots x_1x_2\cdots x_n) \\
&=& f_{(n),*}(x_1x_2\cdots x_n)
f_{(n),*}(x_1x_2\cdots x_n)\cdots
f_{(n),*}(x_1x_2\cdots x_n)
\end{array}\]
\[f_{(kn),*}(x^k)=(f_{(n),*}(x))^k\]
holds for all $x=x_1x_2\cdots x_n\in Q^n$.
\end{proof}
\begin{lemma}
Let $k$ and $n$ be positive integers.
If $CA{-}|f|_{*-*}((2k+1)n)$ is reversible, then
so is $CA{-}|f|_{*-*}(n)$.
\end{lemma}
\begin{proof}
 The result easily follows
from a fact that the identity
\[\begin{array}{ccll}
&& f_{((2k+1)n),*-*}((x_1x_2\cdots x_nx_nx_{n-1}\cdots x_1)^k x_1x_2\cdots
 x_n)\\
&=& (f_{(n),*-*}(x_1x_2\cdots x_n)
f_{(n),*-*}(x_nx_{n-1}\cdots x_1))^k f_{(n),*-*}(x_1x_2\cdots x_n)
\end{array}\]
holds for all $x_1x_2\cdots x_n\in Q^n$.
\end{proof}


\section{$CA{-}\{204,51\}(n)$}

%
%

Triplet local rules $f_{\{204,51\}}:Q^{3}\to Q$ are given, defined by
$f_{204}(abc)=b$ and $f_{51}(abc)=b^-$ for
all triples $abc\in Q^3$. So it holds that
$f^{-}_{204}=f_{51}$, $[204,204,204,204]$ and
$[51,51,51,51]$. Hence equalities
\[f_{204(n),\{a-b,a-*,*-b,*-*,*\}}=\id_{Q^n}\quad
\mbox{and} \quad f_{51(n),\{a-b,a-*,*-b,*-*,*\}}=c_n\]
hold. Thus all configurations in $CA-204(n)$ are
fixed points,
i.e. $CA-204(n)\cong 2^n\langle 1\rangle$, and all
configurations in $CA-51(n)$ lie on limit cycles
of period $2$,
i.e. $CA-51(n)\cong 2^{n-1}\langle 2\rangle$
where $\langle n \rangle$ denotes a limit cycle of period $n$.
\begin{corollary}
$CA-\{204,51\}_{\{a-b,a-*,*-b,*-*,*\}}(n)$
are reversible for all positive integers $n$.
\end{corollary}


\section{$CA{-}\{240,170,15,85\}(n)$}

%
%
%
%

Since $f_{240}(abc)=a$, $f_{170}(abc)=c$,
$f_{15}(abc)=a^-$ and $f_{85}(abc)=c^-$ for all
triples $abc\in Q^3$, we have $f^-_{240}=f_{15}$,
$[240,170,240,170]$ and $[15,85,15,85]$.
Thus it is trivial that $f_{\{240,170,15,85\}(n),*}$
are bijective for all positive integers $n$.
\begin{corollary}
$CA-\{240,170,15,85\}_*(n)$ are reversible
for all positive integers $n$.
\end{corollary}
\begin{lemma}
\begin{enumerate}
\item $f_{15(n),\{a-b,a-*,*-b,*-*,*\}}
=f_{240(n),\{a^--b,a^--*,*-b,*-*,*\}}\circ c_n$,
\item $f_{85(n),\{a-b,a-*,*-b,*-*,*\}}
=f_{170(n),\{a-b^-,a-*,*-b^-,*-*,*\}}\circ c_n$.
\end{enumerate}
\end{lemma}
\begin{proof}
 We will show only
$f_{15(n),a-b}=f_{240(n),a^--b}\circ c_n$.
\[\begin{array}{ccll}
f_{15(n),a-b}(x_1x_2\cdots x_n)
&=& f_{15}(ax_1x_2)f_{15}(x_1x_2x_3)\cdots
f_{15}(x_{n-1}x_nb) \\
&=& a^-x_1^-x_2^-\cdots x_{n-1}^- \\
&=& f_{240}(a^-x_1^-x_2^-)f_{240}(x_1^-x_2^-x_3^-)
\cdots f_{240}(x_{n-1}^-x_n^-b) \\
&=& f_{240(n),a^--b}(x_1^-x_2^-\cdots x_n^-) \\
&=& (f_{240(n),a^--b}\circ c_n)(x_1x_2\cdots x_n).
\end{array}\]
\end{proof}
Remark. $(f_{240(n),*})^n=(f_{170(n),*})^n=\id_{Q^n}$
and $(f_{15(n),*})^n=(f_{85(n),*})^n=(c_n)^n$.
\begin{lemma}
$CA-\{240,170,15,85\}_{\{a-b,a-*,*-b,*-*\}}(n)$
are not reversible for all positive integers $n$.
\end{lemma}
\begin{proof}
 It simply follows from
\[f_{240(n),\{a-b,a-*,*-b,*-*\}}(a^n)
=f_{240(n),\{a-b,a-*,*-b,*-*\}}(a^{n-1}a^-)=a^n.\]
\end{proof}

\section{$CA{-}\{90,165\}(n)$}

%
%

It is obvious that $f_{90}(abc)=a+c$ and
$f_{165}(abc)=(a+c)^-$ for all triples $abc\in Q^3$,
and $[90,90,165,165]$.
Hence by Corollary \ref{corollary-symmetric-reverse} we have
\[CA{-}165_{\{a-b,a-*,*-b,*-*,*\}}(n)\cong
CA{-}90_{\{a^--b^-,a^--*,*-b^-,*-*,*\}}(n)\]
and so we will inspect only $CA{-}90(n)$.
\begin{lemma} \label{CA-90-basic}
 Let $x=x_1x_2\cdots x_n\in Q^n$.
 If $f_{90(n),\{a-b,a-*,*-b,*-*,*\}}(x)=0^n$,
 then $x_i=x_{i+2}$ for all $i=1,2,\cdots,n-2$.
\end{lemma}
\begin{proof}
Set $f=f_{90}$. The condition
$f_{90(n),\{a-b,a-*,*-b,*-*,*\}}(x)=0^n$ means
\[f(*\,x_1x_2)f(x_1x_2x_3)\cdots
f(x_{n-2}x_{n-1}x_n)f(x_{n-1}x_n\,*)=0^n.\]
Hence we have
$x_i+x_{i+2}=f(x_ix_{i+1}x_{i+2})=0$
for all $i=1,2,\cdots,n-2$.
\end{proof}

\begin{lemma}\label{lemma-90-fixed}
\begin{enumerate}
 \item $CA{-}90_{a-b}(n)$ is reversible iff so is
       $CA{-}90_{0-0}(n)$.
 \item $CA{-}90_{0-0}(n)$ is reversible iff $n=0~(\bmod~2)$.
\end{enumerate}
\end{lemma}

\begin{proof}
\begin{enumerate}
 \item 
It is immediate from a fact that
$f_{90(n),a-b}(x)=f_{90(n),0-0}(x)+a0^{n-2}b$ for
all $x\in Q^n$.
 \item 
       First we will show that $f_{90(n),0-0}$
       is injective for $n=0~(\bmod~2)$, since
       $f_{90(n),0-0}$ is additive.
       (Cf. Lemma \ref{fact-additive}.) \\
       (i) Set $f=f_{90}$. It holds that
       $f_{(2),0-0}(x_1x_2)=f(0x_1x_2)f(x_1x_20)
       =x_2x_1$. Hence $CA{-}90_{0-0}(2)$ is reversible.\\
       (ii) Assume that $CA{-}90_{0-0}(n)$
       is reversible for $n\ge 2$, i.e. $f_{(n),0-0}$
       is injective. We will see that $f_{(n+2),0-0}$
       is also injective.\\
       Assume
       $f_{(n+2),0-0}(x_1x_2\cdots x_nx_{n+1}
       x_{n+2})=0^{n+2}$.
       Then we have
       \[
       \begin{array}{ccll}
	f_{(n),0-0}(x_1x_2\cdots x_n)
	 &=& f(0x_1x_2)f(x_1x_2x_3)\cdots f(x_{n-1}x_n0) \\
	&=& 0^{n-1}f(x_{n+1}x_{n+2}0) \\
	&=& 0^n,
       \end{array}
       \]
       since $x_{n-1}=x_{n+1}$ and $x_n=x_{n+2}$ by
       Lemma \ref{CA-90-basic}. Hence by the induction
       hypothesis we have $x_1x_2\cdots x_n=0^n$ and
       consequently $x_{n+1}=x_{n+2}=0$. Therefore
       $CA{-}90_{0-0}(n+2)$ is reversible. 
       Finally we see that if $n=1~(\bmod~2)$ then
       $f_{90(n),0-0}$ is not injective.
       This follows at once from a fact that
       \[f_{90(2k-1),0-0}((10)^{k-1}1)=0^{2k-1}\]
       holds for all positive integers $k$.
\end{enumerate} 
\end{proof}

\begin{corollary}
$CA{-}90_{a-b}(n)$ is reversible iff $n=0~(\bmod~2)$.
\end{corollary}

%
\noindent
In the same discussion as Lemma \ref{lemma-90-fixed}
the following lemma can be shown. 
\begin{lemma}
\begin{enumerate}
 \item $CA{-}90_{a-*}(n)$ is reversible iff so is $CA{-}90_{0-*}(n)$.
 \item $CA{-}90_{0-*}(n)$ is reversible for all positive integers $n$.
\end{enumerate}
\end{lemma}
%
%
%

\begin{corollary}
$CA{-}90_{\{a-*,*-b\}}(n)$ are reversible for all
positive integers $n$.
\end{corollary}

\begin{lemma}
$CA{-}90_{\{*-*,*\}}(n)$ are not reversible for all
positive integers $n$.
\end{lemma}
\begin{proof}
 
It directly follows from
$f_{90(n),\{*-*,*\}}(1^n)=f_{90(n),\{*-*,*\}}(0^n)=0^n$.
\end{proof}

\section{$CA{-}\{102,60,153,195\}(n)$}

%
%
%
%
%

It is obvious that $f_{102}(abc)=b+c$, $f_{60}(abc)=a+b$,
$f_{153}(abc)=(b+c)^-$, $f_{195}(abc)=(a+b)^-$
for all triples $abc\in Q^3$, and $[102,60,153,195]$.
Hence by Corollary \ref{corollary-symmetric-reverse} cellular automata
$CA{-}\{60,153,195\}(n)$ are isomorphic to $CA{-}102(n)$
and so we will inspect only $CA{-}102(n)$.

\begin{lemma}\label{CA-102-basic}
Let $x=x_1x_2\cdots x_n\in Q^n$.
If $f_{102(n),\{a-b,a-*,*-b,*-*,*\}}(x)=0^n$,
then $x_1=x_2=\cdots=x_{n-1}=x_n$.
\end{lemma}
\begin{proof}
 Set $f=f_{102}$. The condition
 $f_{(n),\{a-b,a-*,*-b,*-*,*\}}(x)=0^n$ means that
 \[f(*\,x_1x_2)f(x_1x_2x_3)\cdots
 f(x_{n-2}x_{n-1}x_n)f(x_{n-1}x_n\,*)=0^n.\]
 Hence we have
 $x_i+x_{i+1}=f(x_{i-1}x_ix_{i+1})=0$
 for all $i=1,2,\cdots,n-1$.
\end{proof}

\begin{lemma}
\begin{enumerate}
 \item $CA{-}102_{\{a-b,*-b\}}(n)$ is reversible iff so is
$CA{-}102_{\{0-0,*-0\}}(n)$.
 \item $CA{-}102_{\{0-0,*-0\}}(n)$ are reversible
for all positive integers $n$.
\end{enumerate}
\end{lemma}
\begin{proof}
\begin{enumerate}
 \item 
 It simply follows from a fact that
$f_{102(n),\{a-b,*-b\}}(x)
=f_{102(n),\{0-0,*-0\}}(x)+0^{n-1}b$ holds.
 \item 
Set $f=f_{102}$. Since $f_{(n),\{0-0,*-0\}}
:Q^n\to Q^n$ is additive (modulo $2$), we will
show that $f_{(n),\{0-0,*-0\}}(x)=0^n$ implies
$x=0^n$ in $CA{-}102_{\{0-0,*-0\}}(n)$ for all
positive integers $n$. 

\noindent
(i) It holds that
$f_{(1),\{0-0,*-0\}}(x_1)=f(*x_10)=x_1$ and
$f_{(2),\{0-0,*-0\}}(x_1x_2)=f(*x_1x_2)f(x_1x_20)
=(x_1+x_2)x_2$. Hence $CA{-}102_{\{0-0,*-0\}}(n)$
is reversible for $n=1,2$. 

\noindent
(ii) Assume that $CA{-}102_{\{0-0,*-0\}}(n)$
for $n\ge 2$ is reversible, i.e.
$f_{(n),\{0-0,*-0\}}(x_1\cdots x_n)=0^n$ implies
$x_1\cdots x_n=0^n$. We now see that
\[f_{(n+1),\{0-0,*-0\}}(x_1\cdots x_nx_{n+1})=0^{n+1}
\mbox{ implies } x_1\cdots x_nx_{n+1}=0^{n+1}.\]
Assume
$f_{(n+1),\{0-0,*-0\}}(x_1\cdots x_nx_{n+1})=0^{n+1}$.
Then we have
\[\begin{array}{ccll}
f_{(n),\{0-0,*-0\}}(x_1\cdots x_n)
&=& f(*x_1x_2)f(x_1x_2x_3)\cdots f(x_{n-1}x_n0) \\
&=& 0^{n-1}f(x_nx_{n+1}0) \\
&=& 0^n,
\end{array}\]
since $x_{n-1}x_n=x_nx_{n+1}$ by Lemma \ref{CA-102-basic}.
Hence by the induction hypothesis we have
$x_1\cdots x_n=0^n$. Therefore
$CA{-}102_{\{0-0,*-0\}}(n+1)$ is reversible.
\end{enumerate}
\end{proof}
\begin{corollary}
 $CA{-}102_{\{a-b,*-b\}}(n)$ are reversible
for all positive integers $n$.
\end{corollary}
\begin{lemma}
$CA{-}102_{\{a-*,*-*,*\}}(n)$ are not reversible
for all integers $n$.
\end{lemma}
\begin{proof}
 It is direct from $f_{102(n),\{a-*,*-*,*\}}(1^n)
 =f_{102(n),\{a-*,*-*,*\}}(0^n)=0^n$. 
\end{proof}

\section{$CA{-}\{150,105\}(n)$}

%
%

It is obvious that
$f_{150}(abc)=a+b+c,~f_{105}(abc)=(a+b+c)^-$ for all triples 
$abc\in Q^3$, 
$f_{150}^-=f_{105}$, $[150,150,150,150]$ and
$[105,105,105,105]$.
Hence by Lemma \ref{fact-complement} the reversibility
of $CA{-}105(n)$ and $CA{-}150(n)$ are equivalent
and so we will inspect only $CA{-}150(n)$.

\begin{lemma}\label{CA-150-basic}
Let $x=x_1x_2\cdots x_n\in Q^n$.
If $f_{150(n),\{a-b,a-*,*-b,*-*,*\}}(x)=0^n$, then
$x_i=x_{i+3}$ for all $i=1,2,\cdots,n-3$.
\end{lemma}
\begin{proof} Set $f=f_{150}$. The condition
$f_{(n),\{a-b,a-*,*-b,*-*,*\}}(x)=0^n$ means
\[f(*\,x_1x_2)f(x_1x_2x_3)\cdots
f(x_{n-2}x_{n-1}x_n)f(x_{n-1}x_n\,*)=0^n.\]
Hence we have
\[x_i+x_{i+3}=f(x_ix_{i+1}x_{i+2})
+f(x_{i+1}x_{i+2}x_{i+3})=0+0=0\]
for all $i=1,2,\cdots,n-3$. 
\end{proof}

%

\begin{corollary}\label{lemma-150-free}
$CA{-}150_{\{*-*,*\}}(n)$ are reversible iff
$n\ne 0~(\bmod~3)$.
\end{corollary}
\begin{proof}
 Set $f=f_{150}$. First we will show that
$f_{(n),\{*-*,*\}}(x)=0^n$ implies $x=0^n$
in $CA{-}150_*(n)$ for $n\ne 0~(\bmod~3)$. \\
(i) It holds that $f_{(1),\{*-*,*\}}(x_1)=x_1$,
$f_{(2),*-*}(x_1x_2)=x_2x_1$,
$f_{(2),*}(x_1x_2)=x_1x_2$,
$f_{(4),*-*}(x_1x_2x_3x_4)=x_2(x_1+x_2+x_3)
(x_2+x_3+x_4)x_3$ and
$(f_{(4),*})^2(x_1x_2x_3x_4)=x_1x_2x_3x_4$.
Hence $CA{-}150_{\{*-*,*\}}(n)$ is reversible for
$n=1,2,4$.\\
(ii) Assume that $CA{-}150_{\{*-*,*\}}(n)$
is reversible for $n\ge 2$, i.e. $f_{(n),\{*-*,*\}}$
is injective. We will see that $f_{(n+3),\{*-*,*\}}$
is injective. \\
 Assume
$f_{(n+3),\{*-*,*\}}
(x_1x_2\cdots x_nx_{n+1}x_{n+2}x_{n+3})=0^{n+3}.$
Then we have
\[\begin{array}{ccll}
f_{(n),\{*-*,*\}}(x_1x_2\cdots x_n)
&=& f(\{x_1,x_n\}\,x_1x_2)f(x_1x_2x_3)\cdots
f(x_{n-1}x_n\,\{x_n,x_1\}) \\
&=& f(\{x_1,x_{n+3}\}\,x_1x_2)0^{n-2}
f(x_{n+2}x_{n+3}\,\{x_{n+3},x_1\}) \\
&=& 0^n,
\end{array}\]
since $x_n=x_{n+3}$ and $x_{n-1}=x_{n+2}$
by Lemma \ref{CA-150-basic}.
Hence by the induction hypothesis we have
$x_1x_2\cdots x_n=0^n$ and so
$x_{n+1}=x_{n+2}=x_{n+3}=0$.
Finally we see that if $n=0~(\bmod~3)$ then
$f_{150(n),\{*-*,*\}}$ is not injective.
It follows at once from a fact that
$f_{150(3k),\{*-*,*\}}((101)^k)=0^{3k}$
holds for all positive integers $k$.
\end{proof}
%

\begin{lemma}
 \begin{enumerate}
  \item $CA{-}150_{a-b}(n)$ is reversible iff so is $CA{-}150_{0-0}(n)$.
  \item $CA{-}150_{a-*}(n)$ is reversible iff so is $CA{-}150_{0-*}(n)$.
  \item $CA{-}150_{0-0}(n)$ is reversible iff $n\ne 2~(\bmod~3)$.
  \item $CA{-}150_{0-*}(n)$ is reversible iff $n\ne 1~(\bmod~3)$.
 \end{enumerate}
\end{lemma}
\begin{proof}
 (1,2) It is direct from
 $f_{150(n),a-b}(x)=f_{150(n),0-0}(x)+a0^{n-2}b$ and 
 $f_{150(n),a-*}(x)=f_{150(n),0-*}(x)+a0^{n-1}$.
 (3,4)
 This can be shown in the same way as the proof of Corollary
 \ref{lemma-150-free}.
\end{proof}
\begin{corollary}
\begin{enumerate}
 \item $CA{-}150_{a-b}(n)$ is reversible iff $n\ne 2~(\bmod 3)$.
 \item $CA{-}150_{a-*}(n)$ is reversible iff $n\ne 1~(\bmod 3)$.
\end{enumerate}
\end{corollary}

\section{$CA{-}\{166,180,154,210,89,75,101,45\}(n)$}

It is obvious that
$f_{166}(abc)=(a+1)b+c$, $f_{180}(abc)=a+b(c+1)$,
$f_{154}(abc)=a(b+1)+c$, $f_{210}(abc)=a+(b+1)c$,
$f_{89}(abc)=(a+1)b+c+1$, $f_{75}(abc)=a+b(c+1)+1$,
$f_{101}(abc)=a(b+1)+c+1$, $f_{45}(abc)=a+(b+1)c+1$
for all triples $abc\in Q^3$, $f_{166}^-=f_{89}$,
$[166,180,154,210]$ and $[89,75,101,45]$.
Hence by Corollary \ref{corollary-symmetric-reverse}
and Lemma \ref{fact-complement} the reversibilities of
cellular automata $CA{-}\{166,180,154,210,89,75,101,45\}(n)$
are equivalent and
so we will inspect only $CA{-}166(n)$.\\
We now use the following notation:
\[x_1^{(1)}x_2^{(1)}\cdots x_n^{(1)}=f_{166(n),*}
(x_1x_2\cdots x_n),\]
\[x_1^{(k+1)}x_2^{(k+1)}\cdots x_n^{(k+1)}
=f_{166(n),*}(x_1^{(k)}x_2^{(k)}\cdots x_n^{(k)})\]
for each configuration $x_1x_2\cdots x_n\in Q^n$.
In other words,
\[x_1^{(k)}x_2^{(k)}\cdots x_n^{(k)}
=(f_{166(n),*})^k(x_1x_2\cdots x_n),\]
where $(f_{166(n),*})^k$ is $k$-th composition of
$f_{166(n),*}$. Also a configuration
$x_1x_2\cdots x_n$ in $CA{-}166_*(n)$ is extended
to an infinite configuration
$(x_m)_{m\in\mathbb{Z}}$ such that $x_m=x_{m'}$
if $m=m'~(\bmod~n)$. 

\begin{lemma}\label{fact-ca166}
In $CA{-}166_*(n)$ an identity
\[x_m^{(2^k)}
=(x_{m-2^k}+1)\Pi_{j=1}^{2^k}x_{m-2^k+2j-1}
+x_{m+2^k}\]
holds for all natural numbers $m$ and $k$.
\end{lemma}
\begin{proof}
(i) In the case of $k=0$ an identity
\[x_m^{(1)}=(x_{m-1}+1)x_m+x_{m+1}\]
holds, since $f_{166}(abc)=(a+1)b+c$. \\
(ii) Set $\delta=f_{166(n),*}$. Assume that
the identity holds for a natural number $k$.
Note that
\[\delta^{2^{k+1}}(x)
=(\delta^{2^k}\delta^{2^k})(x)
=\delta^{2^k}(x_1^{(2^k)}x_2^{(2^k)}\cdots
x_n^{(2^k)}).\]
Hence by using the induction hypothesis twice
we have
\[\begin{array}{ccll}
x_m^{(2^{k+1})}
&=& (x_{m-2^k}^{(2^k)}+1)\Pi_{j=1}^{2^k}
x_{m-2^k+2j-1}^{(2^k)}+x_{m+2^k}^{(2^k)} \\
%
&=& \{(x_{m-2^{k+1}}+1)\Pi_{j=1}^{2^k}
x_{m-2^{k+1}+2j-1}+x_m+1\} \\
&& \Pi_{j=1}^{2^k}\{(x_{m-2^{k+1}+2j-1}+1)
\Pi_{i=1}^{2^k}x_{m-2^{k+1}+2j+2i-2}+x_{m+2j-1}\} \\
&& +(x_m+1)\Pi_{j=1}^{2^k}x_{m+2j-1}+x_{m+2^{k+1}} \\
\\
&& \{~m-2^{k+1}+2j+2i-2=m \mbox{ for } i=2^k-j+1~\} \\
\\
&=& \{(x_{m-2^{k+1}}+1)\Pi_{j=1}^{2^k}x_{m-2^{k+1}+2j-1}
+x_m+1\}\Pi_{j=1}^{2^k}x_{m+2j-1} \\
&& +(x_m+1)\Pi_{j=1}^{2^k}x_{m+2j-1}+x_{m+2^{k+1}} \\
&=& (x_{m-2^{k+1}}+1)\Pi_{j=1}^{2^{k+1}}
x_{m-2^{k+1}+2j-1}+x_{m+2^{k+1}}, \\
\end{array}\]
which completes the proof.
\end{proof}

\begin{corollary}
$CA{-}166_*(n)$ is reversible iff $n=1$ ($\bmod~2$)
\end{corollary}
\begin{proof}
 It is trivial that $f_{166(1),*}(x_1)
=x_1$ and so $f_{166(1),*}$ is bijective.
Next we will see that every transition
function  $f_{166(2n-1),*}:Q^{2n-1}\to Q^{2n-1}$
of $CA{-}166_*(2n-1)$ is bijective for all integers
$n\ge 2$. Take a unique integer $k$ such that
$2^k<2n-1<2^{k+1}$.
By the virtue of the last lemma \ref{fact-ca166}
the $m$-th state of $(f_{166(2n-1),*})^{2^k}(x)$
is given by
\[x_m^{(2^k)}
=(x_{m-2^k}+1)\Pi_{j=1}^{2^k}x_{m-2^k+2j-1}+x_{m+2^k}\]
Set $j=n$. (Remark $2\le n\le 2^k$.)
Then $m-2^k+2j-1=m-2^k~(\bmod 2n-1)$ and so
$(x_{m-2^k}+1)x_{m-2^k+2j-1}=(x_{m-2^k}+1)x_{m-2^k}=0$.
Hence we have
\[x_m^{(2^k)}=x_{m+2^k},\]
which proves the bijectivity of $f_{166(2n-1),*}$.
Finally we see that every transition function
$f_{166(2n),*}$ is not injective. It follows
at once from
\[f_{166(2n),*}((01)^n)=f_{166(2n),*}((10)^n)=0^{2n}.\]
This completes the proof.
\end{proof}

\begin{lemma}
 \begin{enumerate}
	\item $CA{-}166_{\{1-b,0-*,*-b\}}(n)$ are not reversible for all
				 positive integers $n$.
	\item $CA{-}166_{\{0-b,1-*,*-*\}}(n)$ are not reversible
				for any positive integers $n\geq 2$.
 \end{enumerate}

\end{lemma}
\begin{proof} It is direct from the following equations;
\begin{itemize}
 \item $f_{166(n),0-b}(110^{n-2})=f_{166(n),0-b}(0^n)$ for $n\geq 2$
 \item $f_{166(1),0-*}(x_1)=0$
 \item $f_{166(2),0-*}(10)=f_{166(2),0-*}(01)$
 \item $f_{166(n),0-*}(110^{n-2})=f_{166(n),0-*}(0^n)$ for $n\geq 3$
 \item $f_{166(1),1-b}(x_1)=b$ 
 \item $f_{166(n),\{1-b,1-*\}}(10^{n-1})=f_{166(n),\{1-b,1-*\}}(0^n)$
       for $n\geq 2$
 \item $f_{166(n),*-b}(10^{n-1})=f_{166(n),*-b}(0^{n})$
 \item $f_{166(n),*-*}(10^{n-1})=f_{166(n),*-*}(0^{n})$ for $n\geq 2$
\end{itemize}
\end{proof}

\section{Conclusion}
We have proved that several 1d CA with finite cell array,
including cyclic CA simulated in \cite{inokuchi-2005}, are
reversible. 
According to computer simulation we can simply observe that 
$CA-R(n)$ except for those which are verified in the paper
are not reversible.
Wolfram showed that for 1d CA with infinite
cell array there exist the only six reversible CA with trivial triplet
local rules.
However the paper presented some nontrivial 1d CA with finite
cell array.
The reversible CA dealt with in this paper are special type
of QCA. 
Our future work is to investigate how these reversible $CA-R(n)$
are extended to generalised PQCA introduced by \cite{inokuchi-2005}.

\end{document}